\documentclass[aps,twocolumn,preprintnumbers,superscriptaddress,floatfix,prl,10pt]{revtex4-2}
\usepackage{bbm}
\usepackage{bm}
\usepackage{amsmath}
\usepackage{amssymb}
\usepackage{empheq}
\usepackage{graphicx}
\usepackage{mathrsfs}
\usepackage{amsfonts}
\usepackage{amsthm}
\usepackage{color}
\usepackage{multirow}
\usepackage{bigints}
\usepackage{txfonts}
\usepackage{float}
\usepackage{hyperref}
\hypersetup{
     unicode=false,              % non-Latin characters in Acrobat?s bookmarks
     pdftoolbar=true,           % show Acrobat?s toolbar?
     pdfmenubar=true,         % show Acrobat?s menu?
     pdffitwindow=false,      % window fit to page when opened
     pdfstartview={FitH},    % fits the width of the page to the window
     pdftitle={My title},       % title
     pdfauthor={Author},     % author
     pdfsubject={Subject},   % subject of the document
     pdfcreator={Creator},   % creator of the document
     pdfproducer={Producer}, % producer of the document
     pdfkeywords={keyword1} {key2} {key3}, % list of keywords
     pdfnewwindow=true, % links in new PDF window
     colorlinks=false,       % false: boxed links; true: colored links
     linkcolor=red,           % color of internal links (change box color with linkbordercolor)
     citecolor=green,        % color of links to bibliography
     filecolor=magenta,    % color of file links
     urlcolor=cyan           % color of external links
}

\makeatletter \tolerance = 10000 \tolerance = 10000

\makeatother
\begin{document}

\title{Linear level repulsions near exceptional points of non-Hermitian systems}

\author{C. Wang}
\email[Corresponding author: ]{physcwang@tju.edu.cn}
\affiliation{Center for Joint Quantum Studies and Department of Physics, School of Science, Tianjin University, Tianjin 300350, China}
\author{X. R. Wang}
\email[Corresponding author: ]{phxwan@ust.hk}
\affiliation{Physics Department, The Hong Kong University of Science and Technology (HKUST), Clear Water Bay, Kowloon, Hong Kong}
\affiliation{HKUST Shenzhen Research Institute, Shenzhen 518057, China}

\date{\today}

\begin{abstract}

The nearest-neighbor level-spacing distributions are a fundamental quantity of disordered systems and are classified into different universality classes. They are the Wigner-Dyson and the Poisson functions for extended and localized states in Hermitian systems, respectively. The distributions follow the Ginibre functions for the non-Hermitian systems whose eigenvalues are complex and away from exceptional points (EPs). However, the level-spacing distributions of disordered non-Hermitian systems near EPs are still unknown, and a corresponding random matrix theory is absent. Here, we show a new class of universal level-spacing distributions in the vicinity of EPs of non-Hermitian Hamiltonians. Two distribution functions, $P_{\text{SP}}(s)$ for the symmetry-preserved phase and $P_{\text{SB}}(s)$ for the symmetry-broken phase, are needed to describe the nearest-neighbor level-spacing distributions near EPs. Surprisingly, both $P_{\text{SP}}(s)$ and $P_{\text{SB}}(s)$ are proportional to $s$ for small $s$, or linear level repulsions, in contrast to cubic level repulsions of the Ginibre ensembles. For disordered non-Hermitian tight-binding Hamiltonians, $P_{\text{SP}}(s)$ and $P_{\text{SB}}(s)$ can be well described by a surmise $\tilde{P}_{\text{ep}}(s)=\tilde{c}_1s\exp[-\tilde{c}_2s^{\tilde{\alpha}}]$ in the thermodynamic limit (infinite systems) with a constant $\tilde{\alpha}$ that depends on the localization nature of states at EPs rather than the dimensionality of non-Hermitian systems and the order of EPs.

\end{abstract}

\maketitle

Symmetries are powerful concepts for classifying disordered quantum systems described by random Hermitian matrices. The nearest-neighbor level-spacing distribution of a disordered metal follows one of three well-known Wigner-Dyson distributions, called symmetry classes, according to time-reversal and spin-rotational symmetries~\cite{wigner_1959,dyson_series,mehta_rmt}. Later, Altland and Zirnbauer proved that the Wigner-Dyson classes, which are invariant by adding a constant potential, do not exhaust all possibilities~\cite{aaltland_prb_1997}. The Wigner-Dyson classes can be further subdivided into seven new groups according to chiral and particle-hole symmetries: three chiral ensembles with chiral symmetry and four Bogoliubov-de Gennes ensembles with particle-hole symmetry. In total, there are ten symmetry classes for Hermitian random matrices.
\par 

Each symmetry class has its specific energy-spectral statistics and features, which are independent of the details of Hamiltonians~\cite{fhaake_book}. Energy-spectral statistics have been studied in many fields of physics, including nuclear physics~\cite{tabrody_rmp_1981}, condensate-matter physics~\cite{cwjbeenakker_rmp_1997}, information theory~\cite{amtulino_book_2004}, and many fundamental phenomena in quantum physics~\cite{fborgonovi_rmp_2016}. One example is the Anderson localization transitions. The distribution $P(s)$ of level spacing $s$ of two nearest-neighbor extended states is well described by the Wigner-Dyson functions of different symmetry classes~\cite{fevers_rmp_2008}. In contrast, $P(s)$ for localized states follows the Poisson distribution. Another example is that energy-spectral statistics can distinguish integrable quantum systems from chaotic ones: the Poisson distribution for quantum integrable systems~\cite{mvberry_prsa_1977} and the Wigner-Dyson distributions for quantum chaotic systems~\cite{obohigas_prl_1984}.
\par 

Non-Hermiticity has a unique position in physics, especially in disordered~\cite{xluo_prl_2021,cwang_prb_2020,nhatano_prl_1996} and topological systems~\cite{atlee_prl_2016,fkkunst_prl_2018,syao_prl_2018}. Level spacing $s$ between two complex eigenenergies is defined as the Euclidean distance in the complex-energy plane such that $P(s)$ is properly defined. A pioneering work by Grobe, Haake, and Sommers shows that $P(s)$ is the Poisson distribution in the complex-energy plane for an integrable system and are the so-called Ginibre distributions of corresponding symmetry classes~\cite{jginibre_jmp_1965} for a fully chaotic system~\cite{rgrobe_prl_1988}. The three Gaussian Ginibre (orthogonal, unitary, and symplectic) ensembles display a universally cubic level repulsion~\cite{rgrobe_prl_1989}, $\lim_{s\to 0}P(s)\sim s^3$, while non-Ginibre distributions also appear in some symmetry classes with transpose symmetry~\cite{gakemann_prl_2019,rhamazaki_prr_2020}. 
\par

Within Ginibre's framework~\cite{jginibre_jmp_1965}, the eigenstates of Hamiltonians are non-orthogonal, and their eigenvalues are generally complex. Nevertheless, a large class of non-Hermitian Hamiltonians possesses exceptional points (EPs) and exceptional lines that separate domains of real eigenenergies from that of complex ones if either parity-time symmetry ($\mathcal{PT}$-symmetry)~\cite{bender_prl_1998} or pseudo-Hermiticity~\cite{amostafazadeh_jmp_2002} is presented. $P(s)$ near EPs, where right eigenstates are mutually orthogonal and their duals are the corresponding left eigenstates, may lead to different energy-spectral statistics than those of Gaussian Ginibre ensembles. However, no careful study of level statistics near EPs is available, and a rigorous extension of random matrix theory (RMT) for EPs is needed.
\par

Our goal is to investigate $P(s)$ near EPs of non-Hermitian systems. We find that the Ginibre distributions are no longer applicable there. The nearest-neighbor level-spacing distributions of small random matrices with EPs, denoted as $P_{\text{ep}}(s)$, are different in the symmetry-preserved and symmetry-broken phases where eigenvalues are real and complex, respectively. Secondly and importantly, level repulsions are {\it linear} near EPs, instead of {\it cubic} in the Ginibre distributions, irrespective of symmetries of non-Hermitian matrices. Thirdly, in the thermodynamic limit, $P_{\text{ep}}(s)$ in both the symmetry-preserved and symmetry-broken phases agree with a surmise of $\tilde{P}_{\text{ep}}(s)=\tilde{c}_1s\exp[-\tilde{c}_2s^{\tilde{\alpha}}]$ with $\tilde{c}_{1,2}$ being normalized constants and $\tilde{\alpha}=2$ and 3 if the state at the EP is extended and localized, respectively. Our surmise $\tilde{P}_{\text{ep}}(s)$ is applicable to a large family of disordered non-Hermitian systems with different orders of EPs and dimensions.
\par

\emph{Symmetry classes with EPs}.$-$We first need to find symmetry classes with EPs. There are eight classes of non-Hermitian Hamiltonians according to four possible symmetry operators $\mathcal{O}$ satisfying $[H,\mathcal{O}]_{\zeta=\pm 1}=H\mathcal{O}-\zeta \mathcal{O}H=0$, where $\mathcal{O}$ are K, Q, P, or C symmetry transformations in the Bernard-LeClair classification. The four allowed transformations are beyond antiunitary and unitary operators required by Hermitian Hamiltonian~\cite{dbernard_book_2002}. Out of the eight non-Hermitian classes, only three of them support real spectra where $\mathcal{O}$ is antilinear, see an analysis in Supplementary Information~\cite{supp}. 
\par

The first two classes are non-Hermitian Hamiltonians with K symmetry, defined by $[H,\Theta_k]_{\zeta=1}=0$, where $\Theta_k=U_k\mathcal{K}$ consists of complex-conjugate operation $\mathcal{K}$ and unitary operator$U_k$. Eigenvalues $\epsilon$ of such a $H$ are either real $\mathbb{R}$ or appear in pairs $(\epsilon,\epsilon^\ast)$, and the critical points separating real and complex eigenvalues are EPs. $\Theta^2_k=\pm I$ distinguish two K-symmetric class with EPs. Here, $I$ is the unit matrix. The eigenstates of a K-symmetric system with $\Theta^2_k=-I$ must be double degenerated, see a proof in Supplementary Information~\cite{supp}. 
\par

The third class is Q-symmetric (also known as pseudo-Hermitian) Hamiltonians satisfying $[H,\Theta_q]_{\zeta=1}=0$, here $\Theta_q=U_q\eta$ is the product of a unitary operator $U_q$ and Hermitian-conjugate operator $\eta$, $\eta\eta^\dagger=\eta^\dagger\eta=I$ and $\eta=\eta^\dagger$~\cite{amostafazadeh_jmp_2002}. One should not confuse the Hermitian-conjugate operator $\eta$ with the complex-conjugate operation $\mathcal{K}$. There is only one Q-symmetric class since $\Theta^2_q=I$, and Hermitian Hamiltonians belong to the trivial Q-symmetric class for $U_q=I$. The remaining five classes featured by P and C symmetries do not imply real spectra and are not considered in this work.
\par

\emph{Small random matrices}.$-$Let us first follow Wigner's wisdom to analytically derive $P_{\text{ep}}(s)$ for small random matrices~\cite{wigner_1959}. We concentrate on Gaussian ensembles whose probability functions are $P(H)dH\propto\exp[-\text{Tr}[HH^\dagger]/\sigma^2]dH$ with $\sigma$ being a real positive number. Consider non-Hermitian Hamiltonians with K symmetry of $\Theta^2_k=I$ and for a specific choice of $\Theta_k=\sigma_1\mathcal{K}$, a $2\times 2$ random matrix with the designed symmetry can be constructed as
\begin{equation}
\begin{gathered}
H^{\text{K},+}_{\text{small}}=a I+b\sigma_1+c\sigma_2+id\sigma_3,
\end{gathered}\label{eq1}
\end{equation}
where $\sigma_{1,2,3}$ are Pauli matrices and $a,b,c,d$ are independent real random numbers with Gaussian distributions of zero means and variance $\sigma^2$. Eigenvalues of $H^{\text{K},+}_{\text{small}}$ are $\epsilon^1_{\pm}=a\pm\sqrt{b^2+c^2-d^2}$, which are real if $b^2+c^2\geq d^2$ and appear in pair, $(\epsilon,\epsilon^\ast)$, if $b^2+c^2< d^2$. The domain with real eigenvalues is termed as the symmetry-preserved phase, and the others known as the symmetry-broken phase~\cite{hyang_prl_2018}. The two phases are separated by an EP at $b^2+c^2-d^2=0$.
\par 

\begin{figure}[htbp]
\centering
\includegraphics[width=0.45\textwidth]{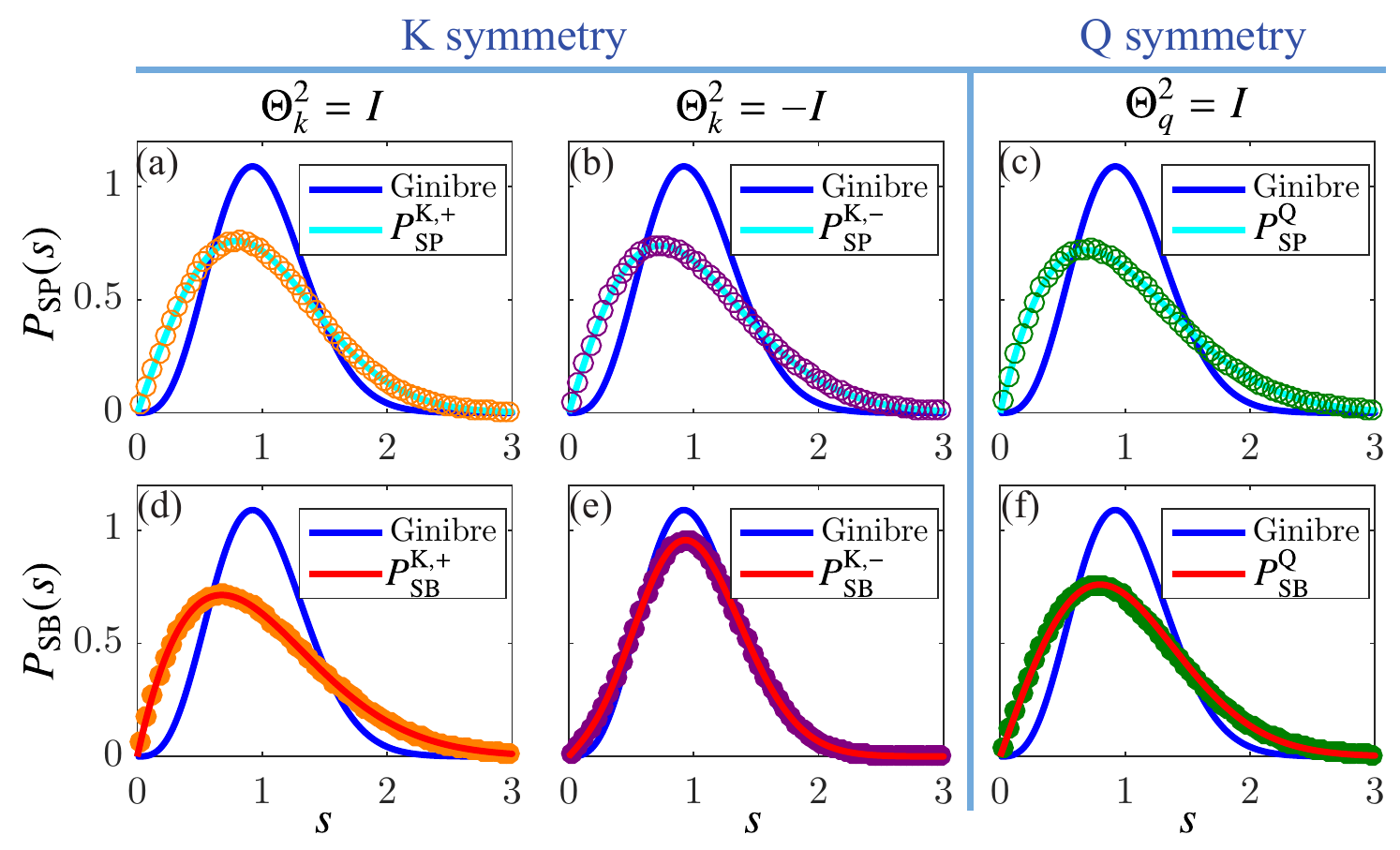}
\caption{The nearest-neighbor level-spacing distributions of small random matrices in the symmetry-preserved (labelled as $P_{\text{SP}}(s)$) and symmetry-broken (labelled as $P_{\text{SB}}(s)$) phases for the K-symmetric classes of $\Theta^2_k=I$ (a,d) and $\Theta^2_k=-I$ (b,e) and the Q-symmetric class (c,f). Empty and filled circles are numerical data obtained by diagonalizing Eqs.~\eqref{eq1},~\eqref{eq3},~\eqref{eq5} for $\sigma=1$ and $10^{10}$ random ensembles, and solid lines are Eqs.~\eqref{eq2},~\eqref{eq4},~\eqref{eq6}. For comparisons, $P(s)$ of the Ginibre unitary distributions of $2\times 2$ matrices are also plotted (blue lines)~\cite{rgrobe_prl_1988}.}
\label{fig1}
\end{figure}

Clearly, $\epsilon^1_{\pm}$ are closest at the EP whose level-spacing distributions are $P_{\text{ep}}(s)$. Since the term inside the square root of $\epsilon^1_{\pm}$ changes signs at the EP, $P_{\text{ep}}(s)$ should be determined by separately integrating over $a,b,c,d$ in the symmetry-preserved and symmetry-broken phases because of different constraints. Let us consider the symmetry-preserved phase first and redefine $b=t\sin[\phi],c=t\cos[\phi],t=\in [0,\infty),\phi\in [0, 2\pi)$. In the symmetry-preserved phase, $\epsilon^1_{\pm}=a\pm\sqrt{t^2-d^2}$ and $t^2>d^2$. Conservation of probability requires 
\begin{equation}
\begin{gathered}
P(a,b,c,|d|)dadbdcd|d|=P(\epsilon^1_+,\epsilon^1_-,t,\phi)\mathcal{J}d\epsilon^1_+ d\epsilon^1_- dt d\phi,
\end{gathered}\label{eq_new1}
\end{equation}
where $\mathcal{J}=(\epsilon^1_+-\epsilon^1_-)t/(2\sqrt{4t^2-(\epsilon^1_+-\epsilon^1_-)^2})$ is the Jacobian. Then, we have
\begin{equation}
\begin{gathered}
P(\epsilon^1_+,\epsilon^1_-)=\int^{2\pi}_0 d\phi \int^{\infty}_{(\epsilon^1_+-\epsilon^1_-)/2}dt P(\epsilon^1_+,\epsilon^1_-,t,\phi)\mathcal{J}\\
=\dfrac{1}{\mathcal{Z}}(\epsilon^1_+-\epsilon^1_-)e^{-[(\epsilon^1_+ + \epsilon^1_-)^2+(\epsilon^1_+ - \epsilon^1_-)^2]/\sigma^2}
\end{gathered}\label{eq_new2}
\end{equation}
with $\mathcal{Z}$ being the normalized constant to be determined. Then, we set $u=\epsilon^1_++\epsilon^1_-$ and $s=\epsilon^1_+ - \epsilon^1_-$ and obtain  $P_{\text{SP}}^{\text{K},+}(s)$ by integrating over $u$ and applying the normalization conditions $\int^{\infty}_0 P_{\text{SP}}^{\text{K},+}(s) ds=\int^{\infty}_0 sP_{\text{SP}}^{\text{K},+}(s) ds=1$:
\begin{equation}
\begin{gathered}
P_{\text{SP}}^{\text{K},+}(s) = (\pi/2)s\exp[-\pi s^2/4]  
\end{gathered}\label{eq_new3}
\end{equation}
Through the same approach, we find $P_{\text{ep}}(s)$ for the symmetry-broken phase is~\cite{supp}
\begin{equation}
\begin{gathered}
P_{\text{SB}}^{\text{K},+}(s) = c_1 s\text{Erfc}[\sqrt{2c_2}s]\exp[c_2 s^2]
\end{gathered}\label{eq2}
\end{equation}
with $\text{Erfc}[x]=(2/\sqrt{\pi})\int^{\infty}_x e^{-t^2}dt$ being the complementary error function ($\lim_{x\to 0}\text{Erfc}[x]=1$) and $c_1\simeq 2.54,c_2\simeq 0.526$.
\par

Equations~\eqref{eq_new3} and \eqref{eq2} accord perfectly with numerical results obtained by directly diagonalizing Eq.~\eqref{eq1}, see Figs.~\ref{fig1}(a) and (d), as well as those for different choices of $\Theta_k$, see evidence in Supplementary Information~\cite{supp}. From Eqs.~\eqref{eq_new3} and \eqref{eq2}, we find $P_{\text{ep}}(s)$ of the two phases exhibit linear level repulsions: $\lim_{s\to 0}P_{\text{SP(SB)}}^{\text{K},+}(s)\sim s$. To the best of our knowledge, linear level repulsions of non-Hermitian random matrices have never been reported before, and the well-known Ginibre distributions predict a cubic level repulsions, $\lim_{s\to 0}P(s)\sim s^3$~\cite{jginibre_jmp_1965,rgrobe_prl_1989}. 
\par

Cubic level repulsions are universal in the Ginibre distributions~\cite{rgrobe_prl_1989}. Naturally, the universality of the linear level repulsions should be tested. Recall that there are two additional classes supporting EPs. The first ones are K-symmetric systems of $\Theta^2_k=-I$, where a two-fold degeneracy is required to obtain EPs~\cite{supp}. Hence, the minimal model is a $4\times 4$ matrix that can be constructed as
\begin{equation}
\begin{gathered}
H^{\text{K},-}_{\text{small}}=aI+ib\Gamma^1+c\Gamma^2+id\Gamma^3+ie\Gamma^4+if\Gamma^5,
\end{gathered}\label{eq3}
\end{equation}  
where $a,b,c,d,e,f$ are independent real random numbers with the same Gaussian distributions. The five anticommuted Gamma matrices are $\Gamma^{1,2,3,4,5}=(I\otimes\tau_3,I\otimes\tau_1,\sigma_1\otimes\tau_2,\sigma_2\otimes\tau_2,\sigma_3\otimes\tau_2)$ with $\tau_{1,2,3}$ being Pauli matrices. One can see that $H^{\text{K},-}_{\text{small}}$ preserves K symmetry since $[H^{\text{K},-}_{\text{small}},\Theta_k]_{\zeta=1}=0$ with $\Theta_k=(i\sigma_2\otimes\tau_1)\mathcal{K}$ and $\Theta^2_k=-I$. Eigenvalues of $H^{\text{K},-}_{\text{small}}$ are doubly degenerated: $\epsilon^2_{\pm}=a\pm\sqrt{c^2-b^2-d^2-e^2-f^2}$. The two degenerated eigenvalues $\epsilon^{2}_{\pm}$ coalesce at an EP where $c^2=b^2+d^2+e^2+f^2$. Analytically, we find $P_{\text{ep}}(s)$ in the symmetry-preserved and symmetry-broken phases are
\begin{equation}
\begin{gathered}
\begin{array}{ccc}
P_{\text{SP}}^{\text{K},-}(s)&=&c_3 \left(\dfrac{s^2e^{-c_4 s^2}}{ c_4}+\dfrac{\sqrt{\pi}s\text{Erfc}[\sqrt{2 c_4}s](1-4c_4 s^2)e^{c_4 s^2}}{(\sqrt{2c_4})^3}\right), \\
\quad\\
P_{\text{SB}}^{\text{K},-}(s)&=&c_5 s(1+4c_6 s^2)e^{-c_6 s^2},
\end{array}
\end{gathered}\label{eq4}
\end{equation}
with $c_3\simeq 1.35,c_4\simeq 0.600,c_5\simeq 0.616,c_6\simeq 1.54$, as well as linear level repulsions $\lim_{s\to 0}P_{\text{SP(SB)}}^{\text{K},-}(s)\sim s$. As shown in Figs.~\ref{fig1}(b) and (e) and Supplemental Information~\cite{supp}, Eq.~\eqref{eq4} accords perfectly with numerical results and is valid for a different $\Theta_k$.
\par

The third symmetry class with EPs is the Q-symmetric class where $[H,\Theta_q]_{\zeta=1}=0$. For simplicity, we choose a specific symmetry operator $\Theta_q=\sigma_3\eta$ such that the corresponding random matrix reads 
\begin{equation}
\begin{gathered}
H^{\text{Q}}_{\text{small}}=aI+ic\sigma_1+id\sigma_2+b\sigma_3,
\end{gathered}\label{eq5}
\end{equation}
where $a,b,c,d$ are the same as those in $H^{\text{K},+}_{\text{small}}$. The eigenvalues are $\epsilon^3_{\pm}=a\pm\sqrt{b^2-c^2-d^2}$. $H^{\text{Q}}_{\text{small}}$ undergoes a transition from the symmetry-preserved phase to the symmetry-broken phase at an EP, $b^2-c^2-d^2=0$, where $P_{\text{ep}}(s)$ in the two phases are derived analytically~\cite{supp}
\begin{equation}
\begin{gathered}
\begin{array}{ccc}
P_{\text{SP}}^{\text{Q}}(s)&=& c_7 s\text{Erfc}[\sqrt{3c_8}s]e^{c_8 s^2}, \\
\quad \\
P_{\text{SB}}^{\text{Q}}(s)&=&(\pi/2)s e^{-\pi s^2/4},
\end{array}
\end{gathered}\label{eq6}
\end{equation} 
with $c_7\simeq 2.42,c_8\simeq 0.271$. Again, we have a linear level repulsion, $\lim_{s\to 0}P_{\text{SP(SB)}}^{\text{Q}}(s)\sim s$, and Eq.~\eqref{eq6} describes numerical data excellently as shown in Figs.~\ref{fig1}(c) and (f). 
\par

Results of small random matrices are simple and meaningful. Although $P_{\text{ep}}(s)$ bifurcate into the symmetry-preserved and symmetry-broken phases and are quantitatively different for different symmetry classes, the level repulsions are always linear. It is widely believed that RMT-statistics lead to cubic level repulsions in non-Hermitian systems, and one would expect that RMT gives cubic level repulsions for all non-Hermitian random Hamiltonians~\cite{rgrobe_prl_1988}. However, Eqs.~\eqref{eq_new3},~\eqref{eq2},~\eqref{eq4}, and~\eqref{eq6} indicate that cubic level repulsions are not true at least near EPs.  
\par

\emph{Large random matrices}.$-$While Wigner-Dyson distributions for small matrices (known as Wigner surmises) are good approximations for random $N\times N$ matrices with $N\gg 1$~\cite{cwang_prb_2017}, Ginibre distributions show significant $N$-dependences~\cite{fhaake_book}. Hence, it is important to investigate $P_{\text{ep}}(s)$ and whether linear level repulsions holds for large matrices near EPs. To calculate $P_{\text{ep}}(s)$, one needs to accurately know EPs. This is easy for small random matrices because analytical expression of eigenvalues are available, but is highly non-trivial for large random matrices in general~\cite{mailybaev_arxiv_2005}. Thus, we consider three special Hamiltonians with K symmetry and with known EPs.
\par

The first one is a tight-binding model in two-dimensional (2D) square lattices of size $L\times L$ whose Hamiltonian in the momentum space and in the absence of disorders is
\begin{equation}
\begin{gathered}
h_{\text{2D}}(\bm{k})=v_0I+\alpha\sin k_1\sigma_2-\alpha\sin k_2\sigma_1+i\kappa\sigma_3
\end{gathered}\label{eq7}
\end{equation}
with $v_0,\alpha,\kappa$ being real positive numbers. The effective $\bm{k}\cdot\bm{p}$ Hamiltonian of Eq.~\eqref{eq7} near $\bm{k}=0$ reads $v_0 I+\alpha(\bm{p}\times\bm{\sigma})\cdot\hat{z}+i\kappa\sigma_3$. The second term describes a Rashba-like spin-orbit coupling with strength $\alpha$~\cite{erashba_spss_1960}, the third term is an imaginary Zeeman term $i\kappa\sigma_3$ distinguishing lifetimes of two orbitals~\cite{vkozii_arxiv_2017}. Possible physical realizations of Eq.~\eqref{eq7} include a large family of ferromagnetic semiconductors such as MnGaAs and other III-V host materials~\cite{nnagaosa_rmp_2010}.
\par

\begin{figure}[htbp]
\centering
\includegraphics[width=0.45\textwidth]{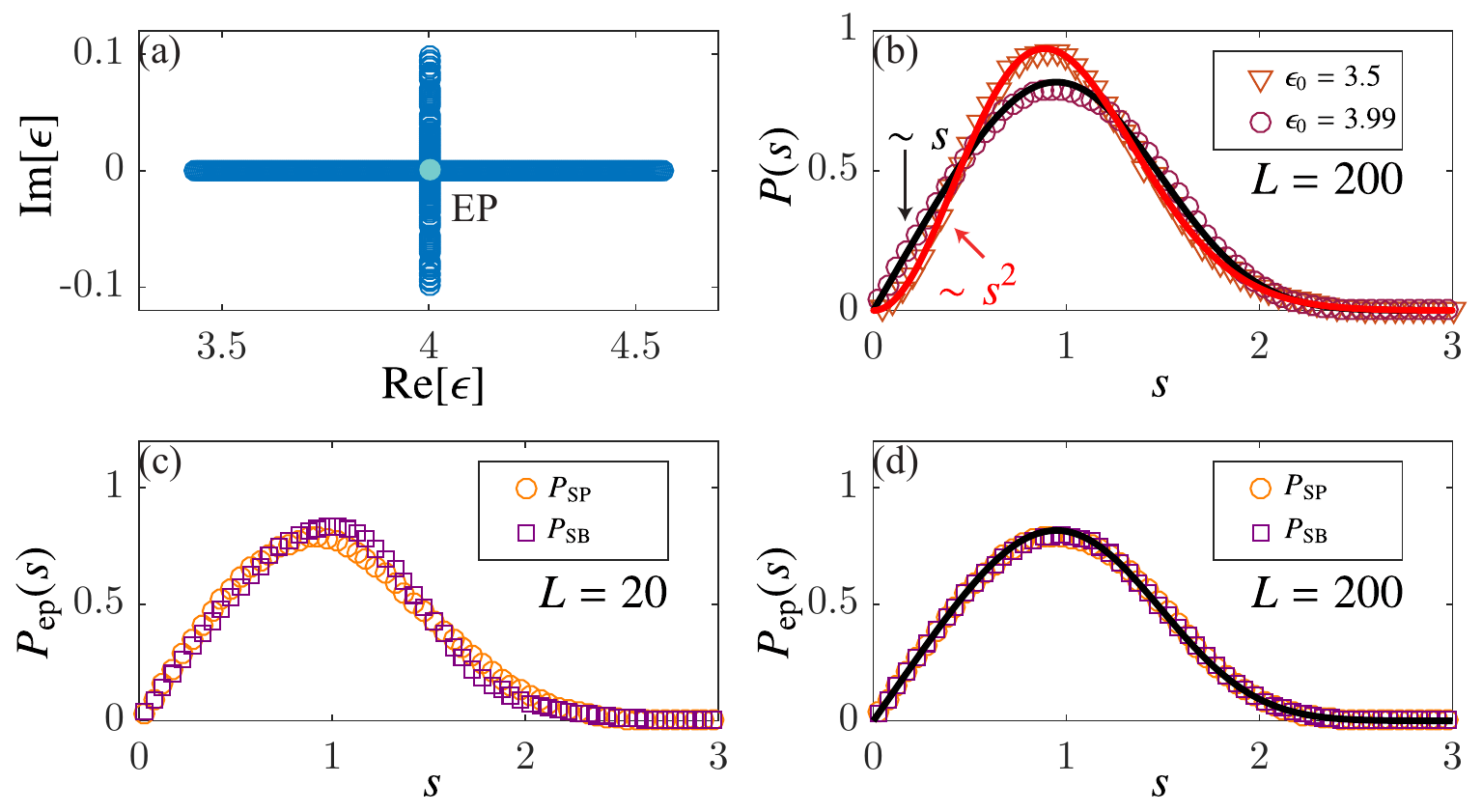}
\caption{(a) Eigenvalues of the real-space Hamiltonian $H_{\text{2D}}$ of Eq.~\eqref{eq7} with disorders in the complex-energy plane for $L=20$. (b) $P(s)$ of $H_{\text{2D}}$ of $L=200$ in two energy windows $[\epsilon_0-\Delta\epsilon,\epsilon_0+\Delta\epsilon]$ with $\epsilon_0=3.5$ (triangles) and 3.99 (circles) and $\Delta\epsilon\sim 10^{-2}$. The red line in (b) is the Wigner surmise for Gaussian unitary ensemble. The black lines in (b) and (d) are $\tilde{P}_{\text{ep}}(s)=\tilde{c}_1 s\exp[-\tilde{c}_2 s^{\tilde{\alpha}}]$ with $\tilde{\alpha}=3$. (c), (d) $P_{\text{SP}}(s)$ and $P_{\text{SB}}(s)$ of $H_{\text{2D}}$ for (c) $L=20$ and (d) $L=200$. Other model parameters are $v_0=4,\alpha=0.2,\kappa=0.1,\sigma=0.1$. Each point in (b)-(d) is averaged over more than $10^4$ ensembles.}
\label{fig2}
\end{figure}

Equation~\eqref{eq7} preserves K symmetry with $\Theta_k=\sigma_1\mathcal{K}$. The disorders are introduced through an on-site random potential $V_{\text{2D}}=\sum_{\bm{i}}c^\dagger_{\bm{i}}v_{\bm{i}}\sigma_2 c_{\bm{i}}$, where $c^\dagger_{\bm{i}}$ ($c_{\bm{i}}$) is  particle creation (annihilation) operator at site $\bm{i}$ and $v_{\bm{i}}$ has a uncorrelated Gaussian distribution of zero mean and variance $\sigma^2$~\cite{supp}. $P_{\text{ep}}(s)$ is obtained by numerically solving $H_{\text{2D}}+V_{\text{2D}}$, where $H_{\text{2D}}$ is Hamiltonian Eq.~\eqref{eq7} in real space whose expression is given in Supplementary Information~\cite{supp}.  Disorders break lattice-translational symmetry but preserve K symmetry. For $\alpha>\kappa/\sqrt{2}$, $N=2L^2$ eigenvalues of $H_{\text{2D}}$ distribute in a cross region in the complex-energy plane with the EP at $\epsilon=v_0+0i$, see Fig.~\ref{fig2}(a). $P_{\text{ep}}(s)$ curves are obtained from two nearest-neighbor eigenvalues to the EP for many random configurations, where the conventional unfolding procedures are used~\cite{tguhr_pr_1998}.
\par

For states in the symmetry-preserved phase far from the EP, say $\epsilon\in[\epsilon_0-\Delta\epsilon,\epsilon_0+\Delta\epsilon]$ with $\epsilon_0=3.5$ and $\Delta\epsilon\sim 10^{-2}$, $P(s)$ in Fig.~\ref{fig2}(b) is well described by the Wigner surmise of Gaussian unitary ensemble [Here, $\lim_{s \to 0} P(s)\sim s^2$]~\cite{cwang_prb_2017}. This is because non-Hermitian systems in the symmetry-preserved phase behave like a Hermitian system without the time-reversal symmetry due to $V_{\text{2D}}$. Near the EP, say $\epsilon_0=3.99$, $P(s)\simeq P_{\text{SP}}(s)$ that deviates from the Wigner-Dyson distribution and shows a linear level repulsion in the limit of $s\to 0$, see Fig.~\ref{fig2}(b). This also happens for $P_{\text{SB}}(s)$ in the symmetry-broken phase. Interestingly, for a small system size $L=20$, $P_{\text{SP}}(s)$ is different from $P_{\text{SB}}(s)$, but they merge for a large system size of $L=200$, see Figs.~\ref{fig2}(c) and (d), respectively.      
\par

Our surmise of the nearest-neighbor level-spacing distributions near the EPs is   
\begin{equation}
\begin{gathered}
\tilde{P}_{\text{ep}}(s)=\tilde{c}_1 s\exp[-\tilde{c}_2s^{\tilde{\alpha}}].
\end{gathered}\label{eq8}
\end{equation}
Here, $\tilde{\alpha}>0$, and $\tilde{c}_{1,2}$ are normalized constants~\cite{constants}. The surmise has the linear level repulsion for small $s$ and an exponential decay $\propto\exp[-\tilde{c}_2 s^{\tilde{\alpha}}]$ for large $s$. For $L=200$, $\tilde{P}_{\text{ep}}(s)$ fits well to the numerically-calculated $P_{\text{SP}}(s)$ and $P_{\text{SB}}(s)$ of Hamiltonian Eq.~\eqref{eq7} with $\tilde{\alpha}=2.99\pm0.02$ and $\tilde{\alpha}=3.02\pm0.03$, respectively, see black lines in Figs.~\ref{fig2}(b) and (d).
\par

\begin{figure}[htbp]
\centering
\includegraphics[width=0.45\textwidth]{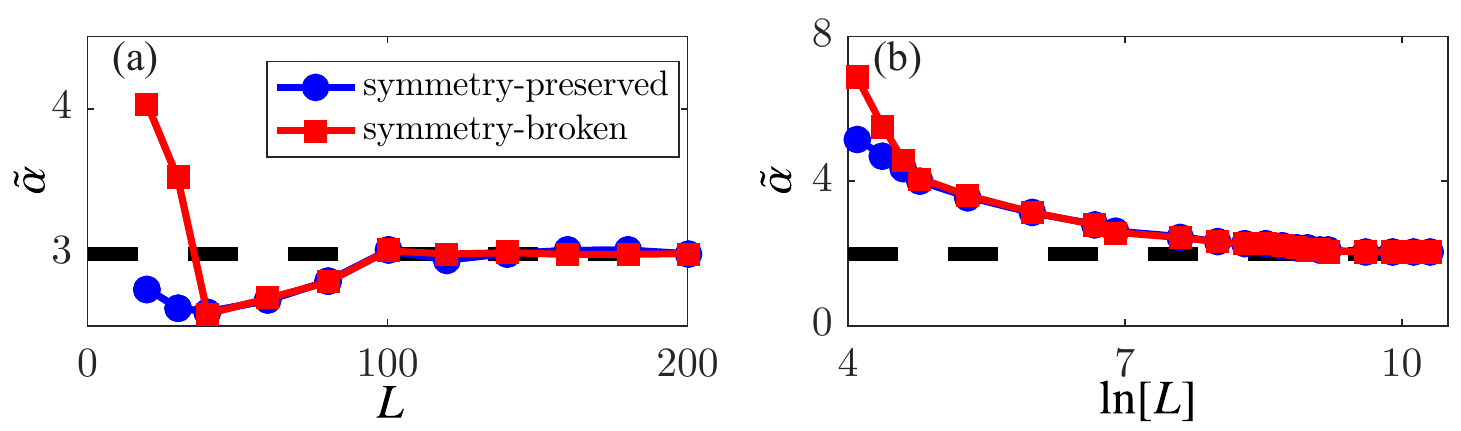}
\caption{(a) $\tilde{\alpha}$ as a function of $L$ for $H_{\text{2D}}$ in the symmetry-preserved (the blue circles) and symmetry-broken (the red squares) phases. (b) $\tilde{\alpha}$ as a function of $\ln[L]$ for  $H_{\text{1D}}$. Here, $\nu_0=4,\alpha=0.2,\kappa=0.1,\sigma=0.1$. The black dashed lines in (a) and (b) locate $\tilde{\alpha}=3$ and 2, respectively.}
\label{fig3}
\end{figure}

$P_{\text{SP}}(s)$ and $P_{\text{SB}}(s)$ for various system sizes $L$ are numerically obtained. The goodness-of-fit $Q$ of our data to Eq.~\eqref{eq8} is $Q>10^{-3}$ for $L>10$ such that Eq.~\eqref{eq8} is a satisfactory description of $P_{\text{ep}}(s)$ for $L>10$. Figure~\ref{fig3}(a) depicts the exponent $\tilde{\alpha}$ as a function of $L$. Similar to small random matrices [$L=1$], $P_{\text{SP}}(s)$ and $P_{\text{SB}}(s)$ for relatively small sizes [say $10<L<40$] are different as $\tilde{\alpha}$ in the symmetry-preserved phase is not equal to that in the symmetry-broken phase. With the increase of $L$, $P_{\text{SP}}(s)$ and $P_{\text{SB}}(s)$ merge and approach $\tilde{P}_{\text{ep}}(s)$ with $\tilde{\alpha}=3$. 
\par

To test whether the exponent $\tilde{\alpha}_{L\to\infty}$ is universal, we consider a one-dimensional (1D) tight-binding model of length $L$ with K symmetry whose Hamiltonian is
\begin{equation}
\begin{gathered}
h_{\text{1D}}(k_1)=v_0 I+\alpha\sin k_1\sigma_2+i\kappa\sigma_3.
\end{gathered}\label{eq9}
\end{equation}
Equation~\eqref{eq9} satisfies $[h_{\text{1D}},\Theta_k]_{\zeta=1}=0$ with $\Theta_k=\sigma_1\mathcal{K}$. The Hamiltonian in real-space is $H_{\text{1D}}$ given in the Supplementary Information. Random on-site potentials $V_{\text{1D}}=\sum_i c^\dagger_i v_i\sigma_2 c_i$ with $v_i$ following the Gaussian distribution of the zero mean and variance $\sigma^2$ are used for studying the level statistics, see Supplemental Information~\cite{supp}. The model has an EP at $\epsilon=\nu_0+i0$. From fitting $P_{\text{ep}}(s)$ to Eq.~\eqref{eq8}, $\tilde{\alpha}=2$, instead of $\tilde{\alpha}_{L\to\infty}= 3$ in 2D, is obtained for the symmetry-preserved and symmetry-broken phases as shown in Fig.~\ref{fig3}(b). Interestingly, $\tilde{\alpha}=2$ equals to the Brody distribution in 2D for independently uniformly distributed random energy levels in the complex-energy plane~\cite{brody_lnc_1973,sa_prx_2020}.
\par

The reason for two $\tilde{\alpha}_{L\to\infty}$ in Fig.~\ref{fig3} is as follows: For Hermitian systems, $P(s)$ at an Anderson transition point universally decays as a stretched-exponential, $\propto\exp[-\tilde{c}_2 s^{\tilde{\alpha}}]$, for large $s$, and becomes a Gaussian ($\tilde{\alpha}=2$) or a Poisson ($\tilde{\alpha}=1$ that is the Brody distribution in 1D) for the extended and localized states, respectively~\cite{cwang_prb_2017}. Based on this fact, we conjecture $P_{\text{ep}}(s)$ for localized EPs follows the Brody distribution in 2D since levels of localized states are uncorrelated. However, $P_{\text{ep}}(s)$ for levels near the extended EPs, which are correlated, has a faster decay rate at the tail, i.e., a larger exponent $\tilde{\alpha}_{L\to\infty}=3$. We have partially confirm this argument by proving the following issues in Supplementary Information~\cite{supp}: (i) EPs of Eq.~\eqref{eq7} undergo an Anderson localization transition at $\sigma_c=0.63\pm 0.05>\sigma=0.1$ used in Fig.~\ref{fig3}(a). (ii) EPs of Eq.~\eqref{eq9} are localized by infinitesimal disorders.
\par

\begin{figure}[htbp]
\centering
\includegraphics[width=0.45\textwidth]{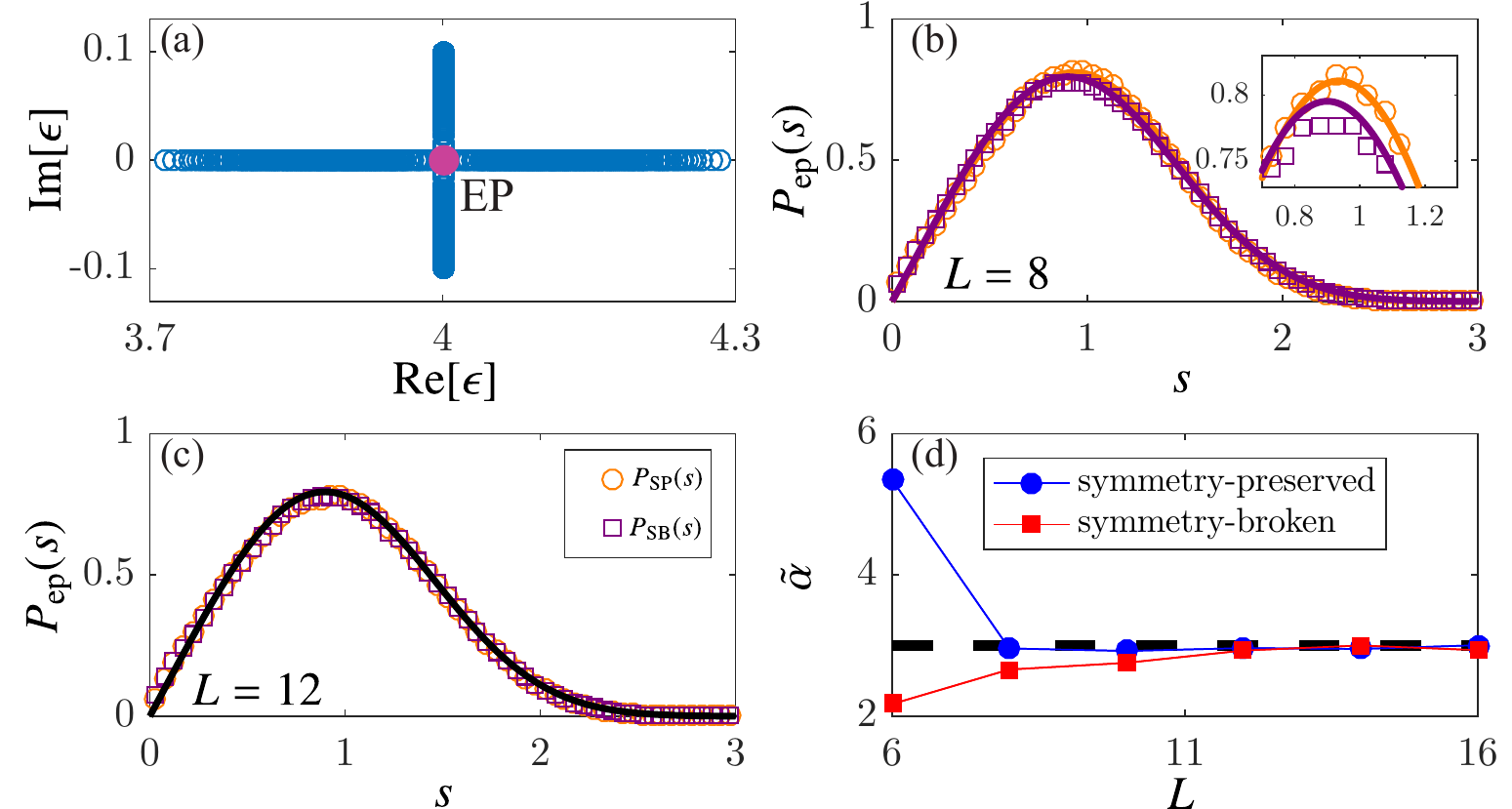}
\caption{(a) Eigenvalues of $H_{\text{3D}}$ of Hamiltonian Eq.~\eqref{eq_new_3} with disorders in the complex-energy plane for $\alpha=0.2,\kappa=0.1,\sigma=0.1,L=8$. (b) $P_{\text{SP}}(s)$ (the orange circles) and $P_{\text{SB}}(s)$ (the purple squares) of levels near the EP in (a). The solid lines are fitted by Eq.~\eqref{eq8} with $\tilde{\alpha}=2.96\pm 0.08$ and $2.6\pm 0.2$ for $P_{\text{SP}}(s)$ and $P_{\text{SB}}(s)$, respectively. Insert: Zoom-in of the peaks of $P_{\text{SP}}(s)$ and $P_{\text{SB}}(s)$. (c) Same as (b) but for $L=12$. The black solid line is Eq.~\eqref{eq8} of $\tilde{\alpha}=3$. (d) $\tilde{\alpha}$ as a function of $L$ for the symmetry-preserved and symmetry-broken phases. The black dashed line is $\tilde{\alpha}=3$. Each data is average over more than $10^4$ ensembles.}
\label{fig4}
\end{figure}

\emph{Higher-order EPs}.$-$EPs in Eqs.~\eqref{eq7} and \eqref{eq9} are second-order. It is important to check whether level-spacing distributions near a higher-order EP exhibit also the linear level repulsion. For this purpose, we consider the following three-dimensional (3D) model of size $L\times L\times L$ whose clean Hamiltonian in the momentum space is 
\begin{equation}
\begin{gathered}
h_{\text{3D}}(\bm{k})=\nu_0 I+\alpha\sum_{\mu=1,2,3}\sin k_\mu \Gamma^{\mu}+i\kappa\Gamma^4.
\end{gathered}\label{eq_new_3}
\end{equation}
$h_{\text{3D}}(\bm{k})$ has K symmetry since $[h_{\text{3D}}(\bm{k}),\Theta_k]_{\zeta=1}=0$ with $\Theta_k=i\sigma_2\otimes\tau_0\mathcal{K}$. For $\kappa=0$, $h_{\text{3D}}(\bm{k})$ is Hermitian and display a quadruple degeneracy, whereby two doubly degenerate bands touch the other two at high-symmetry points in the first Brillouin zone. For finite $\kappa$, the degeneracy points split into forth-order EPs at $\alpha^2(\sin^2 k_1+\sin^2 k_2+\sin^2 k_3)=\kappa^2$, see Supplemental Information~\cite{supp}. The forth-order EPs form a closed exceptional sphere of radius $\kappa/\alpha$ in the Brillouin zone.
\par

We study the real-space Hamiltonian $H_{\text{3D}}$ of Eq.~\eqref{eq_new_3} with an additional random on-site potential $V_{\text{3D}}=\sum_{\bm{i}}c^\dagger_{\bm{i}}v_{\bm{i}}\Gamma^1 c_{\bm{i}}$ where $v_{\bm{i}}$ is the Gaussian distribution of 
zero mean~\cite{supp}. The disordered potential does not break K symmetry, and the EP is at $\epsilon=\nu_0+i0$ in the complex-energy plane, see Fig.~\ref{fig4}(a). 
\par

Akin to those of the second-order EPs, $P_{\text{SP}}(s)$ and $P_{\text{SB}}(s)$ of the forth-order EPs of $L=8$ can be fitted by Eq.~\eqref{eq8} with $\tilde{\alpha}=2.96\pm 0.08$ and $2.6\pm 0.2$, respectively, see Fig.~\ref{fig4}(b). It is about $\tilde{\alpha}=3$ for a larger size $L=12$ as shown in Fig.~\ref{fig4}(c). Furthermore, as shown in Fig.~\ref{fig4}(d), $\tilde{\alpha}$ of $P_{\text{SP}}(s)$ and $P_{\text{SB}}(s)$ merge and approach to 3 in the thermodynamic limit, similar to the cases of $h_{\text{2D}}$ shown in Fig.~\ref{fig3}(a). 
\par

Generally speaking, states in 3D models are much more extended than those in 2D models. We have proven that the EPs of $H_{\text{2D}}$ of the same disorder strength $\sigma=0.1$ are extended. It is reasonable to assert that the EPs of $H_{\text{3D}}$ are extended as well. Hence, Fig.~\ref{fig3}(a) and Fig.~\ref{fig4}(d) strongly indicate that the order of EPs and the dimensionality of non-Hermitian systems do not change $P_{\text{ep}}(s)$ where $\lim_{s\to 0}P_{\text{ep}}(s) \sim s$ and $\tilde{\alpha}=3$, as long as the EPs are extended. 
\par

\emph{Discussions}.$-$With the rapid advance in Hamiltonian engineering in optical~\cite{aguo_prl_2009}, mechanical~\cite{cmbender_ajp_2013}, electric~\cite{sassawaworrarit_nature_2017} systems, to name a few, where EPs are realized by suitably controlling gain and loss, the reported linear level repulsion can be tested experimentally. Here, we suggest cavity-magnon-polaritons as feasible platforms for observing linear level repulsions at EPs, whose effective Hamiltonians are non-Hermitian due to the inevitable loss. The $\mathcal{PT}$-symmetric systems with EPs have already been realized experimentally~\cite{dzhang_nc_2017,mharder_prl_2018,wyu_prl_2019,jdubail_pra_2022}, and quasi-particles due to strong couplings between magnons and cavity photons were detected. Our prediction should be easily detectable in this well-developed system, see Supplementary Information~\cite{supp}.
\par

\emph{Conclusion}.$-$In summary, the nearest-neighbor level-spacing distributions near EPs display linear level repulsions for small random matrices. We generalize this finding by investigating 1D, 2D, and 3D disordered tight-binding Hamiltonians with either the second-order or the forth-order EPs and find that the profile Eq.~\eqref{eq8} of $P_{\text{ep}}(s)$ describes our numerical data for large enough sizes well. One interesting open question is whether there exist other classes of EPs with nonlinear level repulsions. Non-Hermitian systems have, in total, thirty-eight symmetry classes if multiple symmetries are considered, in which twenty-eight classes support EPs~\cite{dbernard_book_2002}. We speculate that all of them exhibit linear level repulsions, but a comprehensive study of all symmetry classes is needed before making a definite statement about the question. 
\par

\begin{acknowledgments}

This work is supported by the National Key Research and Development Program of China 2020YFA0309600, the National Natural Science Foundation of China (Grants No.~11704061 and No.~11974296), and Hong Kong RGC (Grants Nos.~16301518 16301619, and 16302321). C. W. acknowledges the kindly help from Dr. Weichao Yu concerning the experimental proposal for observing the linear level repulsion.

\end{acknowledgments}


\begin{thebibliography}{99}
\bibitem{constants}
$\tilde{c}_1=(\tilde{\alpha}\Gamma[3/\tilde{\alpha}]^2)/(\Gamma[2/\tilde{\alpha}]^3)$ and $\tilde{c}_2=\left( \Gamma[3/\tilde{\alpha}]/\Gamma[2/\tilde{\alpha}] \right)^{\tilde{\alpha}}$ determined by $\int^\infty_0 \tilde{P}_{\text{ep}}(s)ds=\int^\infty_0 s\tilde{P}_{\text{ep}}(s)ds=1$. Here, $\Gamma[x]$ is the Gamma function.
\end{thebibliography}
\end{document}